\RequirePackage{fix-cm}
\documentclass[smallextended]{svjour3}       
\smartqed  
\usepackage{graphicx}
\usepackage[hyphens]{url}
\usepackage[hyperindex,breaklinks]{hyperref}

\usepackage[sort&compress,numbers]{natbib}
\usepackage{doi}

\hypersetup{
  colorlinks=true,
  linkcolor=blue,
  filecolor=magenta,
  urlcolor=blue,
  citecolor=blue,
}

\begin{document}

\title{Renormalization of $NN$ Chiral EFT
}
\subtitle{my personal mixed feelings}


\author{D.R. Entem}


\institute{D.R. Entem \at
              Grupo de F\'\i sica Nuclear and IUFFyM, \\ Universidad de Salamanca, E-37008 Salamanca, Spain \\
              Tel.: +34-670588539\\
              \email{entem@usal.es}           
}

\date{Received: date / Accepted: date}

\maketitle

\begin{abstract}
We are celebrating thirty-years from the seminal papers from Weinberg proposing to use
the rules of Chiral Perturbation Theory to nucleon systems. His proposal was to build
an Effective Field Theory (EFT) with Chiral symmetry as a key property.
A big effort on this line
have been made during these three decades, however the issue of renormalization of the
theory is not settled. I will briefly review my path on this issue and give my personal
view.
\end{abstract}

\section*{} 

At beginning of the 1990's Weinberg proposed to use the rules of Chiral Perturbation Theory
to nuclear systems~\cite{WEINBERG1990288,WEINBERG19913}. However he noticed that Chiral
Perturbation Theory should breakdown due to the enhancement of $NN$ intermediate states,
and he proposed to use the rules to calculate the potential instead of the scattering amplitude, and then
introduce it in a Schr\"odinger or Lippmann-Schwinger type equation. This allows to
resum the infinite reducible diagrams generated by the iteration of irreducible ones, and
so, being able to generate non-perturbative effects as the well known $NN$ bound state, the deuteron.
This is the same idea used to calculate the hydrogen atom using Quantum Electrodynamics.

At that time I was an undergraduate physics student at University of Salamanca, and very soon I knew
about Weinberg. Indeed, he visited Salamanca in 1992 to received the Honoris
Causa Doctorate from University of Salamanca, although I didn't have the chance to see him.
Very soon after, my studies introduced me in Quantum Field Theory and the concept of
renormalizability that led to the construction of Quantum Chromodynamics.

However, later on, I started my PhD thesis studying the $NN$ interaction in the framework of constituent
quark models and saw how in nuclear systems always cut-offs appear in the interactions.
For me it was a little bit annoying to see that some times results have more sensitivity on these,
in principle, unphysical cut-offs, than in the real physical constants of the model.

However the use of cut-offs in nuclear physics was very common. The reason to use them was that
terms needed to build the interactions do not vanish for infinite momentum, and this was a way
to regulate the Lippmann-Schwinger equation. The way
to understand the appearance of cut-offs is of course that nuclear systems are not made of
point-like particles, and so a form factor could be present.

During the 1990s there was a very important development of Weinberg's ideas and the first
$NN$ potentials based on Chiral EFT were developed~\cite{ORDONEZ1992459,PhysRevLett.72.1982,PhysRevC.53.2086,
KAISER1997758,EPELBAOUM1998107,EPELBAUM2000295}. I started to work in $\chi$EFT
when I went as a postdoc to University of Idaho to work with Ruprecht Machleidt.
At the beginning we were quite skeptical for several reasons~\cite{Machleidt:2021ggx}. 

At that time one of the things that shocked me was that using finite cut-offs in a
QFT was not a problem, I was not used to the EFT concept. 
Also part of the community had the 
impression that $\chi$EFT was doing the same as the old one-boson-exchange models
or the Bonn model.

However, as it is well known, there are many differences. The main one is that there is a
systematic way to build the theory which is basically given by the symmetries and
effective degrees of freedom. This is achieved using a certain power counting. The
first $NN$ power counting introduced by Weinberg was naive dimensional analysis. This
introduce a new perturbative expansion in terms of ratios of scales, in particular
the ratio of low energy external momenta over a hard energy scale of the theory
given by the chiral-symmetry-breaking scale $\Lambda_\chi$, which gives a low
energy expansion of the interaction. 

The power counting is the key of the EFT and there is an important discussion about
which is the correct one~\cite{KAPLAN1998390,PhysRevC.74.014003,PhysRevC.83.024003,PhysRevC.84.064002,PhysRevC.84.057001,PhysRevC.85.034002,PhysRevC.86.024001,fbs.54.2175,epja.41.341,Machleidt_2010,PhysRevC.88.054002,epja.56.152}.
For me one of the important properties is that it 
gives the terms in the Lagrangian that have to be introduced and the diagrams
in the potential to be computed. With a correct power counting all the infinities generated
in irreducible diagrams can be absorb by terms in the Lagrangian, and naive dimensional
analysis accomplish that.
Weinberg's implicit assumption was that also the infinities generated by
reducible diagrams could be absorb by the same terms in the Lagrangian.
However the question is what happens with the non-perturbative resummation of the reducible ones.

The infinities are only generated when loop integrals are done in an infinite momentum space.
There is a way to implement the EFT program considering a scale $\Lambda$ between
the low-energy scales and the hard energy scale $\Lambda_\chi$. The scale $\Lambda$ gives the maximum momentum in the
loop integrals and there is no infinities anymore. This is accomplished cutting the integrals
in loop momentum or introducing some regulator function. However results have to be independent
of this scale (and the regulator function), so renormalization is seeing as independence of
the observables on the regulator.

This is the main idea under the different $NN$ potentials that have been developed and have been
quite successful~\cite{PhysRevC.68.041001,MACHLEIDT20111,EPELBAUM2005362,
PhysRevC.96.024004,Reinert2018}. 
Phenomenologically this prescription works very well~\cite{PhysRevC.88.054002}. The main benefit of
this approach is that once you have a potential you can compute observables for nuclear systems
in ab initio calculations. So from a pragmatic point of view, it is the ideal solution. I learned
that sometimes in physics you need to be pragmatic from Machleidt, my mentor and colleague in $\chi$EFT.

However, always small cut-off artifacts are present which is unpleasant. Also, seeing the big
success of the renormalization program for QED without cut-offs, one would like
to accomplish it even in an EFT. This is the other way to implement the EFT program, which is
the correct one is a matter of discussion~\cite{10.3389/fphy.2020.00079}.

With this point of view I was introduced to the so called non-perturbative renormalization by
E. Ruiz-Arriola. The idea is to implement the same renormalization program when the
Lippmann-Schwinger equation is used. He and M.P. Valderrama were using the so called
renormalization with boundary conditions for $NN$ $\chi$EFT~\cite{PhysRevC.74.054001,PhysRevC.74.064004}, that had also
been used by Nogga et al.~\cite{PhysRevC.72.054006}. We checked that it was
the same as performing the renormalization with a counter term in momentum space~\cite{PhysRevC.77.044006}. This
enhanced my interest in trying to completely remove regularization dependence.

However there was a big problem, for singular repulsive interactions there was a definite
solution with no possible renormalization. For singular attractive interactions only one low
energy constant could be fixed. These constraints made impossible to give a good description
of $NN$ data and even made impossible to have a convergent pattern in the $\chi$EFT expansion~\cite{Zeoli:2012bi}.

In 2013 I attended an ESNT workshop in Paris to discuss, among others, this problem. There,
I met J.A. Oller who presented the results for $NN$ using the N/D method and $\chi$EFT
and we started a collaboration. The goal was to include the non-perturbative
sum of reducible diagrams in the N/D method. We were able to find an integral equation for
the left hand discontinuity (which is the input of the N/D method) that was equivalent to
the Lippmann-Schwinger equation. This is what we called the exact N/D method~\cite{ENTEM2017498,OLLER2019167965}.

The exact N/D method is equivalent to the solution of the Lippmann-Schwinger equation for regular
potentials. However, the interesting point is that it can also be used for singular interactions.
We first checked
if using only one or no subtractions in the N/D method we could obtain the same results
as the previous approaches. The agreement was perfect.

The advantage of the N/D method is that one can introduce more subtractions. In the perturbative case,
when one uses the left hand cut discontinuity of the $T$-matrix for a finite number of diagrams, it was proven that
the solution exists, even for singular terms~\cite{PhysRevC.89.014002}. 
For example, for regular terms any arbitrary number of subtractions can be
made. 

However in the non-perturbative case the situation is different. The integral equation for
the left hand discontinuity is always finite, so there is no need to regulate it. But the
N/D method uses the left hand discontinuity for imaginary on-shell momenta going to infinity and the
equation generates a solution that diverges on this limit. This is not always a problem
for the N/D method as mentioned previously. However in the non-perturbative case, since we don't know
the asymptotic behavior of the potential, we don't have a way to know a priori if a solution exists or not.
We had to rely on numerical solutions and we saw that more subtractions could be made, even in the
singular repulsive case. These were really good news.

However, having a solution doesn't always mean having a meaningful solution. 
In $\chi$EFT we don't
know the real solution, so we didn't know if we could rely on it. However, using an idea introduced by 
Epelbaum {\it et al.}~\cite{Epelbaum_2018}, we studied a potential that was regular if all pieces were included~\cite{EPJST}. 
This defined what we called the solution
of the full theory, since it corresponds to the well defined solution in the regular case. However this potential had long-range, middle-range
and short-range contributions, and they were chosen so that the middle-range were singular. Removing
the short range part we just have the same as in $\chi$EFT, a problem with singular interactions.
Now we could make more than one subtraction, fixing them to the low energy parameters of the full theory,
and compare with the result of the full theory. The results for low energy were very good
when more than one subtraction was used. And the good agreement appear not only in the 
singular attractive case, but in the singular repulsive one~\cite{EPJST}. This solves the problem previously
mention and gives hope to give a good description of $NN$ observables with complete
regularization independence. What is still not clear is if a potential could be build from this
calculation, which could be the price to pay.

So, even though the community (as myself) have not decide which is the correct
way to renormalize $\chi$EFT, using the first approach a proper description
of $NN$ observables can be made as shown by different potentials, and I believe 
that the second approach will also work.

The great impact of Weinberg's proposal in the nuclear community is out of question and
I believe that we have many reasons to celebrate it.

\begin{acknowledgements}
This work has been funded by
Ministerio de Ciencia e Innovaci\'on
under Contract No. PID2019-105439GB-C22/AEI/10.13039/501100011033,
and by EU Horizon 2020 research and innovation program, STRONG-2020 project, under grant agreement No 824093.
\end{acknowledgements}


\bibliographystyle{apsrev4-1}
\bibliography{Entem}

\end{document}